# Clinicopathological correlation of p40/TTF1 co-expression in NSCLC and review of related literature

LIAN YANG, MING XIAO, XIAN LI, YA-LAN WANG
Department of Pathology, Molecular Medicine and Cancer Research Center,
Chongqing Medical University, Chongqing 400016, P.R. China

TTF1 and ΔNp63/p40 have been used to differentiate ADC and SQC in hypofractionated NSCLC because of their sensitivity and specificity. There are few cases where TTF1 and ΔNp63/p40 are expressed together in the same tumour cells, and little is known about the clinicopathological features, treatment and prognosis of such cases. We investigated the electron microscopic features, immunohistochemical expression and molecular variation of a case of TTF1/p40 co-expressing NSCLC and reviewed and summarised the relevant literature. Our patient was a 58-year-old male with a CT showing a left-sided lung occupancy. As in all other cases reported in the literature, the tumour showed a solid growth pattern with polygonal cells, eosinophilic cytoplasm and clearly visible nuclear fission. Immunohistochemistry showed positive for TTF-1, p40, CK5/6, CK7, P63 and p53, and negative for NapsinA and ALK. Electron microscopy showed tumour cells characterised by bidirectional differentiation of adenocytes and squamous cells, consistent with previous reports. Second-generation sequencing suggested co-mutation of STK11/LKB1 and NF1 genes in this case. Mutations in STK11/LKB1 and NF1 genes have been found in ADC and SQC and are often associated with drug resistance and poor prognosis, but STK11/NF1 co-mutation has not been reported and more cases are needed to reveal the association. p40/TTF1 co-expression in NSCLC may be an under-recognised variant of NSCLC The origin may be a double positive stem cell-like basal cell located in the distal airway, with rapid clinical progression and poor prognosis.

*Correspondence to:* Dr Ya-Lan Wang, Department of Pathology, Molecular Medicine and Cancer Research Center, Chongqing Medical University, Chongqing 400016, P.R. China E-mail: wangyalan@cqmu.edu.cn

## Introduction

Advances in chemotherapy and targeted therapies targeting specific molecular alterations have the need for precise subclassification of nonsmall-cell lung cancer(NSCLC) including adenocarcinoma (ADC) and squamous cell carcinoma (SQC) primarily.Poorly differentiated non-small cell lung cancer (PD-NSCLC) is a predominantly solid growth tumor formed by cells with abundant and eosinophilic cytoplasm, a misleading keratinizing tendency or pseudosquamous cell morphology. Classification is difficult, particularly for small biopsy specimens, and histochemical mucin assays and immunohistochemistry are often required to aid the differential diagnosis and classification of NSCLC.

Positivity for thyroid transcription factor (TTF-1), napsin-A and cytokeratin 7 (CK7) tends to be diagnostic of ADC, whereas positivity for p63, ΔNp63/p40 (hereafter referred to as p40) ,CK5/6, CK34bE12, desmoglein-3 or desmocollin supports the diagnosis of SQC[1-4]. TTF-1 is expressed in the basal cells of the alveolar epithelium and distal bronchiolar epithelium that constituting the terminal respiratory unit (TRU), TTF-1 is considered to be the most specific and sensitive markers for ADCs origin of TRU [5].TTF-1 is a highly specific marker (97–100% specificity) but not very sensitive (54-75% sensitivity), while CK7 is more sensitive (usually >90%) but less sensitive specific marker (specificity 57–94%) [6-7]. P40 is currently considered to be the most specific marker of SQC , with close to 100% sensitivity and specificity to squamous cell differentiated tumour cells [8-9]. Therefore,TTF1/p40 has been widely used as the preferred immunohistochemical combination to distinguish ADC from SQC[8,10-11].Lung tumor tissue microarray (TMA) data showed that the combined use of TTF1 and p40 markers had 93% sensitivity and 92% specificity in the diagnosis of SQC[12]. For the few cases that remain difficult to diagnose, the addition of the second-line antibodies CK5/6 ,P63 and CK7 markers can be considered to help confirm the diagnosis.The phenotype p40+/TTF1- corresponds to SQC, and phenotype p40-/TTF1+ or double negative supports the diagnosis of ADC [10-11].Cases of concurrent P40 and TTF1 positivity on the same tumour cells are very rare , with the exception of the present case, only seven cases have been reported in detail in the literature so far(Table 1)[13-18]. In addition, the occurrence of this rare immunophenotype has been reported in large scale biopsy studies and named as “amphicrine” biphenotypic tumor, but no further studies were carried out[10,19].

We report a case of p40/TTF1 co-expressing NSCLC and compare other reported cases, analysing its clinicopathological features from clinical, immunohistochemical (IHC), electron microscopic (EM) and next-generation sequencing （NGS） perspectives, to suggest possible implications for the classification, diagnosis and treatment of NSCLC .

Table 1. Clinicopathological features of P40/TTF1 co-expressed NSCLC and cases reported in the literature

| Author | Sex/Age(y) | Smoking History | Radiographic Imaging | Tumor metastasis and staging | Histology | IHC | Electron microscopy | Molecular mutation and frequency | Medical Therapy | Clinical Outcome |
|---|---|---|---|---|---|---|---|---|---|---|
| Yang lian (present case) | Male/58 | Current smoker (40 years,360 pack year) | The left upper lobe ( 19mm) | No lymph node metastases, slight pleural thickening | PD- NSCLC with favored squamous differentiation | p40（clone ERB）TTF-1（clone 8G7G3/1）positive in most tumor cell nuclei | The cytoplasm of tumour cells was microvillous and extended, with intracellular mucus granules and perinuclear tension filaments | STK11 (exon5:c.613G>A:p.A205T) 2.5% NF1exon26:c.3394C>T:p.R1132C 2.3% | CAR-NK cell therapy | Alive |
| Pelosi(**2015**)[13] | Male/77 | Former smoker (40 pack-year ) | The left hilar（85mm） | involving major bronchi and associated with ipsilateral pleural effusion **Stage:**c-IVA | PD- NSCLC with hints of squamous differentiation | P40 (clone BC28 ) TTF1 (clone 8G7G3/1) positive in most tumor cell nuclei | Extracellular lumen formation with microvilli-like protrusions; Squamous lineage: mucous granules and abundant perinuclear tonofilaments | K-RAS (AAA>AAT K117N exon4)32%, TP53(GTG>GGG, V272G exon 8)71% | None | Died of respiratory failure a month and a half after the first hospitalization. |
| Hayashi(2018)[14] | Male/73 | Former smoker (141 pack-year ) | The left upper lobe (19mm) | NA | PD-NSCLC with necrotic solid component, focal adenoid differentiation but negative for mucus staining | P40 (clone BC28 ) TTF1 (clone 8G7G3/1) positive in most tumor cell nuclei | NA | PTEN (pHis123Asp)11%, TP53 (pVal272Leu)9.6% | NA | NA |
| Spinelli(**2019**)[15] | Male/ 51 | Current smoker (NA) | The right upper lobe(31mm) | Mediastinal LNs, brain, bone, adrenal glands(stage cT2a N3 M1c | PD- NSCLC with hints of squamous differentiation | P40 (clone RP-163-05 ) TTF1 (clone SPT-24) positive in most tumor cell nuclei | NA | TP53(c.215C>G(p. Pro72Arg) | radiotherapy followed by chemotherapy to brain metastases | Died of disease three months after the initial diagnosis |
| Pelosi(2021)[16] | Female/62 | Former smoker (10 pack-year ) | The right upper lobe(45mm) | Mediastinal LNs,brain（stage pT4N2) | PD- NSCLC with comedo-type necrosis and peripheral palisading | P40 (clone RP-163-05 ) TTF1 (clone SPT-24) positive in most tumor cell nuclei | Intercellular lumina bordered by microvilli and fascicles of keratin fibers | EGFR(p.Glu746_ALA750del, c.2235_2249del)61%; TP53( p.Glu224Asp,c.672G>T)53%; VUS(RAD51B, p.Pro365Arg, c1094C>G)41% ; CCND3( p.Ser259Alac775T>G )47% | Chemotherapy and gefitinib | Died of disease 48 months after the initial diagnosis |
| Pelosi(2021)[16] | Male/62 | Former smoker (23 pack-year ) | The left lower lobe(47mm) | Mediastinal LNs,thorax, liver（stage ypT2bN2) | PD- NSCLC with hints of squamous differentiation | P40 (clone RP-163-05 ) TTF1 (clone SPT-24) positive in most tumor cell nuclei | Intracytoplasmic lumina and fascicles of keratin fibers | NF1( pArg1769Ter, c.5305C>T)54% | Chemotherapy plus pembrolizumab | Died of disease three months after the initial diagnosis |
| Li Hui(2021)[17] | Female/54 | No smoking history | The left upper lobe(NA) | Mediastinal LNs, brain, bone | PD- NSCLC with hints of squamous differentiation | p40（clone ERB）TTF-1（clone 8G7G3/1）positive in most tumor cell nuclei | NA | EGFR(p.L747_S752del,c.2239 _2256del);TP53(p.C135R,c.403 T ＞C) | Chemotherapy and pemetrexed | Alive |
| Chen Bing(2022)[18] | Female/58 | NA | The left upper lobe (22mm) | NA | PD- NSCLC with prominent cellular atypia , visible nucleoli and mitotic figures | P40（NA）TTF-1（NA）positive in most tumor cell nuclei | NA | NA | NA | Losing contact |

## Case Report

The case was a 58-year-old male smoker (360 pack-years). The computed tomography scan showed scattered nodules in both lungs, and the high-risk nodules in the posterior segment of the upper lobe of the left lung were about 1.9 cm in diameter (Fig. 1),with a slightly thickened pleura.In addition, multiple foci of calcification were seen in the lower lobe of the right lung and fibrous bands were seen in lower lobe of the left lung. After admission, thoracoscopic radical surgery for malignancy in the upper lobe of the left lung was performed, and the patient subsequently received anti-infective treatment and CAR-NK cell therapy. The patient was 9 months post-surgery and had no significant discomfort at the time of the return visit.

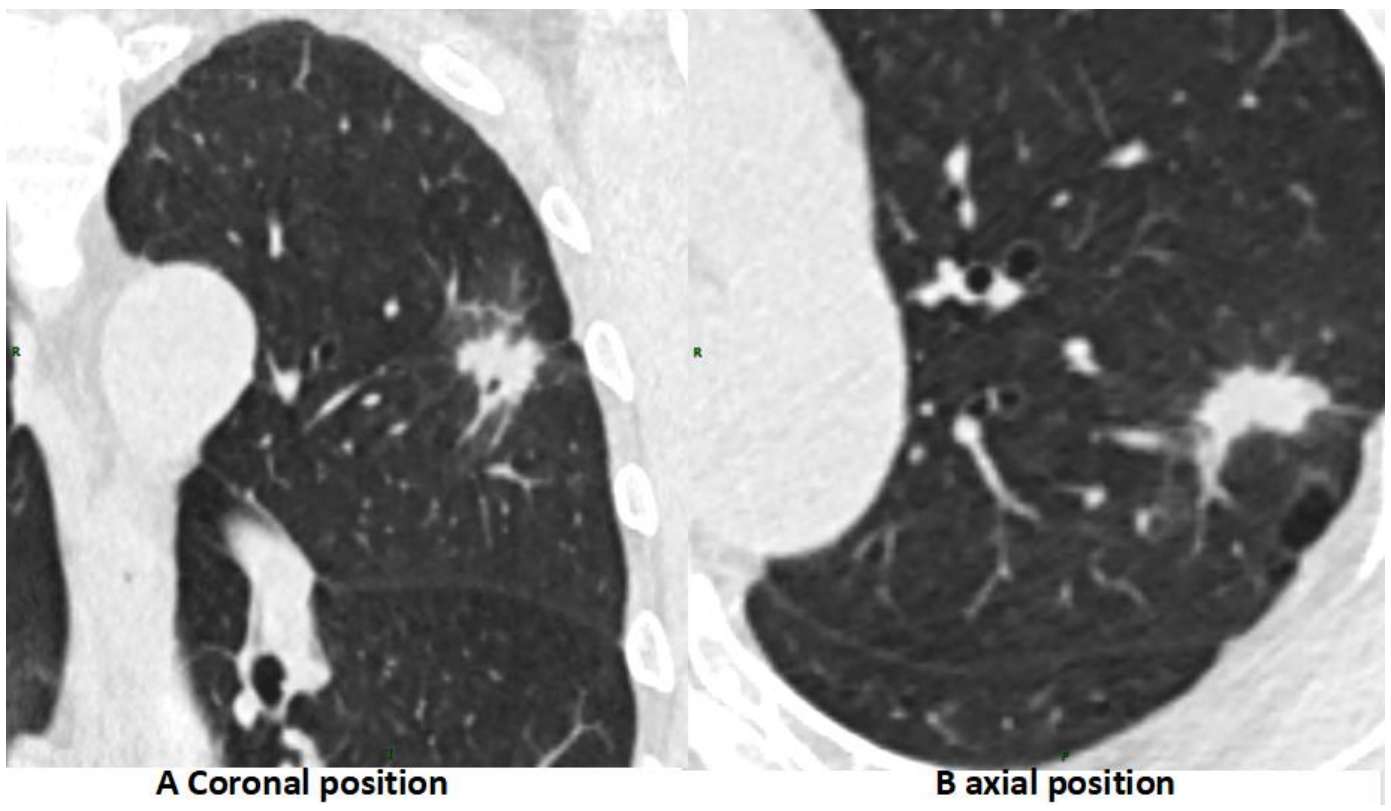

Fig. 1 CT scan :The mass is located in the upper lobe of the left lung and is surrounded by a glassy shadow, measuring approximately 1.9cm x 1.3cm, with lobulated, burred margins and small internal vacuoles.Fibrous bands also be seen in the basal segment of the lower lobe of the left lung.

The surgically resected specimen was a greyish-white, solid mass, partially cystic and poorly demarcated from the surrounding tissue. HE staining of the biopsy tissue showed that the tumour was a poorly differentiated non-small cell carcinoma with a nested or diffuse distribution of cancer cells. The cytoplasm was red eosinophilic and the cancer cells had features that favoured squamous cell differentiation (Figure 2 A and B). Immunohistochemical staining showed extensive and intense positivity for p40 and TTF1 in the nuclei of the same tumour cells (Figure 2 C and D), supporting double-labelled expression in the same tumour cells. There were no lymph node metastases on pathological examination.

Electron microscopic observation of the ultrastructural field showed that the cytoplasm of tumour cells with dual P40 and TTF1 co-expression was microvillous and extended, with a large number of intracellular mucus granules and abundant perinuclear tension filaments (Figure 2 E and F), showing both glandular and squamous cell characteristics.

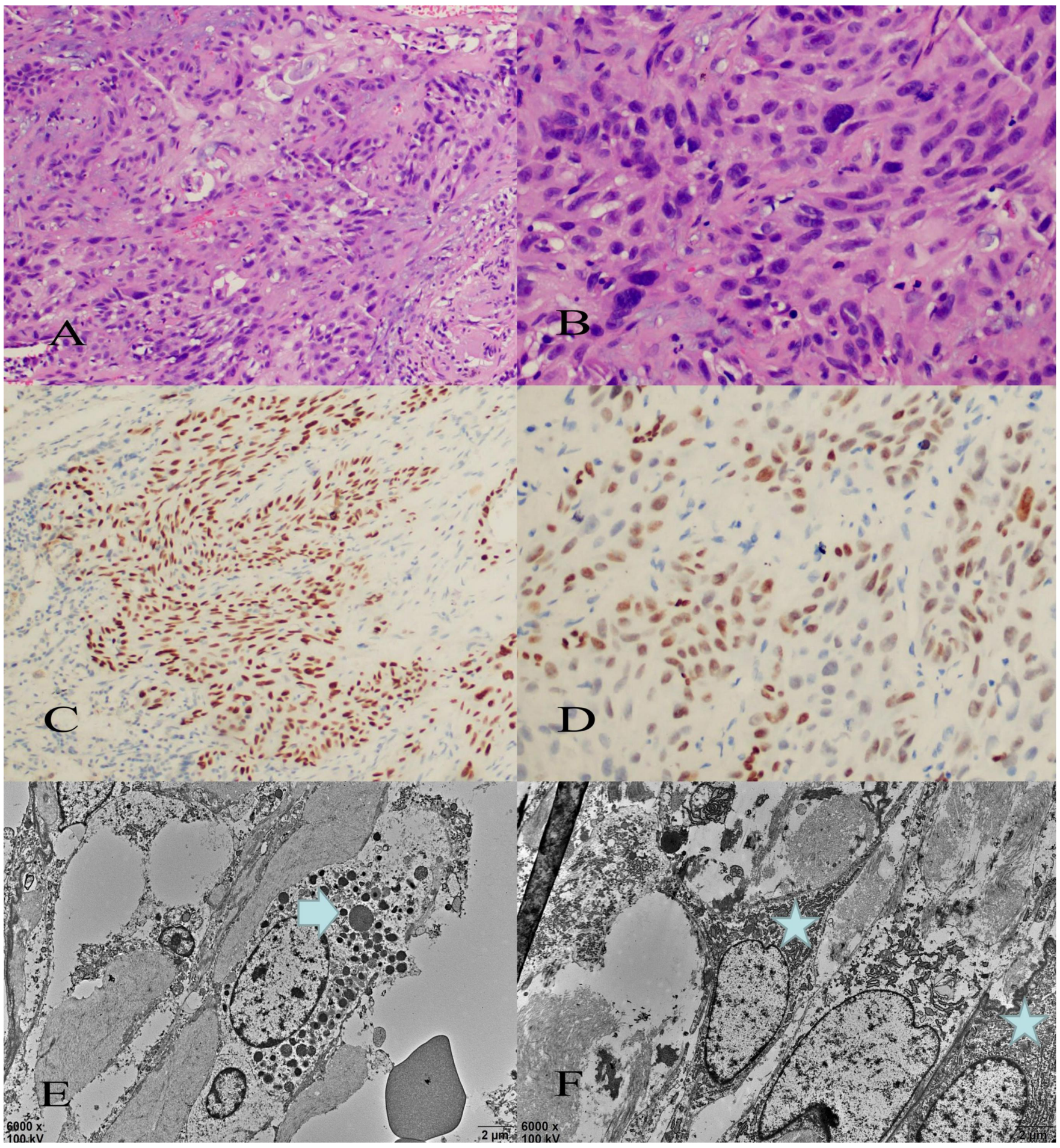

Fig. 2　Histology showed poorly differentiated NSCLC　with a predominantly solid nested component, partly characterised by poorly differentiated squamous carcinoma, with individual cells showing a mucinous cytoplasmic component (A.B). Immunohistochemical staining revealed the presence of both nuclear TTF1 and p40 positive reactions in most of the same tumour cells （C.D）.Electron microscopy shows this dual differentiation of squamous and glandular cells at the same individual cell level. The cytoplasm of the tumour cells is microvillous and extended, with numerous intracellular mucus granules (arrows) and perinuclear tension filaments (asterisks) （E.F）.

Next generation sequencing was performed by ShengTing Medical Laboratory (sample number: 21A011308), The detection method is capture NGS sequencing. The scope of sequencing includes 160 genes and PD-L1 (clone 22C3) protein expression in solid tumour pan-cancer species. Refer to the　Catalogue of Somatic Muation in Cancer database.A total of 8 mutations in 8 genes were detected in this sample, of which 2 mutations were related to targeted drugs, and 6 genes mutations have been reported in the literature and in public databases, but there is no clear evidence of clinical application for association with specific tumours， the clinical significance was temporarily unknown.

Table 2. Detection of somatic gene mutation information by NGS

| Gene | Test results | Mutation frequency | clinical trial drug |
|---|---|---|---|
| NF1 | exon26 c.3394C>T p.R1132C | 2.3% | Selumetinib,trametinib |
| STK11 | exon5 c.613G>A p.A205T | 2.5% | Vistusertib,Nirapali |
| ARID1A | exon16:c.3978_3980del:p.Q1334del | 6.5% | Clinical significance is unknown |
| BRCA2 | exon3:c.310G>A:p.D104N | 6.7% | Clinical significance is unknown |
| FGFR1 | exon5:c.372_374del:p.D125del | 7.0% | Clinical significance is unknown |
| MSH2 | exon1:c.7G>A:p.V3M | 7.6% | Clinical significance is unknown |
| RET | exon17:c.2924A>T:p.N975I | 7.2% | Clinical significance is unknown |
| TSC1 | exon6:c.357_358AG:p.H120D | 7.8% | Clinical significance is unknown |

Notes:

1. Mutations of unknown clinical significance are those that have been reported in the literature and public databases, but for which there is no clear evidence of clinical application in relation to a specific tumour.

2.Refer to the COSMIC(database of oncogene somatic mutations) for classification and criteria for clinical significance of gene variants.

## Discussion

ADC and SQC, the two main subtypes of NSCLC , differ in their histological patterns, immunohistochemistry, key gene mutations and treatment.The morphology of poorly differentiated NSCLC is often not indicative of differentiation and immunohistochemistry will be an important basis for tumour classification. The human p63 gene, located on chromosome 3q27-29, contains 2 promoters and produces 2 types of proteins: the full-length protein TAp63 (containing the N-terminal transcriptional activation domain) and the truncated protein ΔNp63 (N-terminal truncated protein TA63), where TAp63 can be labelled with the antibody p63 and ΔNp63 is named p40 [20]. P40 is commonly found in the basal layer of stratified epithelium and some glandular epithelium, where it has almost 100% sensitivity and specificity in SQC [21]. This may be because P40 has mechanisms to maintain the stem cell state, resist differentiation towards columnar epithelium and induce squamous cell differentiation [20].

TTF-1 is a specific marker for lung ADC. Simultaneous expression of TTF-1 and p63 has been reported in 5.5% of cases [22-23]. Compared with TTF-1/ p63 co-expression, NSCLC with co-expression of TTF1/P40 are rarer. There are only eight cases reported so far, including this case (TABLE 1).From the clinical data, five of the eight cases were male ,with an age range of 51-77 years (median age 60 years) and a clear history of smoking in all but two cases. Six cases were located in the left lobe of the lung and two in the right lobe, with a mean tumour diameter range of 19-85 mm. Five cases had metastases at presentation and three died within three months after treatment, suggesting a poor prognosis for this type of lung cancer.

Histologically almost all cases showed a poorly differentiated (high-grade) NSCLC in pathological morphology, with pleomorphic tumor cells, abundant cytoplasm and varying degrees of squamous differentiation characteristics. All cases characteristically appeared to have both P40 and TTF1 widely and strongly expressed in the nuclei of tumour cells, which is inconsistent with the reported expression pattern of ADC or ADSQC, that both P40 and TTF1 have their respective positive regions [11,24] which may be because when one of the genes is transcriptionally dominant, SQC (under p63) or ADC (under TTF1) differentiation is mutually exclusive[8,10,25].

Electron microscopic examination of our case also showed features of both glandular and squamous differentiation. Glandular differentiation included extracellular lumen formation, villi-like cytoplasmic protrusions and mucus granules; squamous differentiation included abundant perinuclear tension filaments, dense cytoplasmic keratin fibrils and bridging granule junctions. This is consistent with the three cases previously reported by Pelosi G[13,16]. We suggest that such cases may represent carcinomas that differentiate into squamous epithelium while retaining stem cell phenotypic features and may arise from rare co-differentiated double-positive basal stem cells, a novel type of non-small cell carcinoma.

Pelosi G et al. have speculated that co-expression of TTF-1 and p40 in NSCLC cells may be caused by bidirectional expression of stem cells, suggesting that such cases should be included in adenosquamous carcinoma or at least NSCLC with an immunophenotype of adenosquamous carcinoma and proposing an explanatory model in which p40/TTF1 double-positive tumours originate from a subpopulation of bronchial basal cells present in the normal distal airways[13,16]. Hayashi et al [14] reported a case of NSCLC with double expression of TTF1 and p40, accompanied by co-mutations of TP53 and PTEN alleles. This author hypothesized that loss-of-function mutations in PTEN lead to combinatorial activation of the PI3K/AKT pathway, which in turn promotes tumorigenesis. TTF1 may co-transcribe with p40 downstream target genes that regulate the development of this type of tumor, resulting in hypofractionated and multi-phenotypic tumors.

TP53 mutations were detected in five of the eight TTF-1/p40 co-expressing NSCLC cases. Spinelli et al. suggested that this may be a new type of lung cancer, with TP53 mutations representing its more aggressive nature [13-17]. Daniela Cabibi et al[26]. observed that basal reserve cells in the terminal respiratory unit (TRU) were also p40 positive, suggesting that TTF1/p40-positive PD-NSCLC may represent a very rare basal immunophenotypic feature of reserve cells, similar to the 'basal-like' tumours of the breast. They are clinically more aggressive and may be a new and under-recognised tumour, suggesting the name 'basal-type TRU carcinoma' ,and that larger multicentre studies are needed to clarify their clinicopathological and genetic features and their histogenetic association with TRU basal cells .

The genetic variants identified by gene sequencing, including this case, include EGFR, TP53, CCND3, VUS, PTEN, K-RAS, NF1, and STK11. Among them, EGFR, NF1, and TP53 gene alterations are commonly found in ADC, while NF1, TP53, and CCND3 are commonly found in SQC [27-29]. These genetic alterations also seem to reflect the bidirectional differentiation characteristics of tumour cell glands and squamous cells, implying that the p40+/TTF1+ NSCLC subtype may be a different type of stem-based cell originating from the distal airways.

The results of genetic testing showed that there were gene mutations of STK11 and NF1 in this case. STK11 has not been reported in the reported NSCLC with co-expression ofTTF1/p40.STK11 gene, also known as LKB1, a typical tumor suppressor gene which expressed in a variety of tissues, encoding a serine/threonine kinase of the calmodulin family and involved in all these important cellular process,it is the major upstream activator of AMPK[30].

Somatic mutations or loss-of-function alterations in STK11/LKB1 are present in a variety of tumors, including NSCLC, pancreatic cancer,cervical cancer, papillary serous ovarian cancer and breast cancer[31-35]. STK11/LKB1 gene alterations occur in up to 30% of NSCLC cases.Squamous or adenosquamous carcinomas have a high incidence of STK11/LKB1 mutations in the expanded tumor histological repertoire, possibly related to epigenetic reprogramming caused by STK11/LKB1 deletion. Mechanistically, transdifferentiation [G] is mediated by STK11/LKB1 deletion-triggered downregulation of the Polycomb repressive complex 2 (PRC2) subunit EED and remission of PRC2-mediated repression of squamous differentiation genes[36].Loss of heterozygosity in the chromosome where STK11/LKB1 is located can be accompanied by nonsense mutations, point mutations, and deletion variants that can lead to STK11/LKB1 inactivation[31], resulting in more active lung cancer cells, increased metastatic potential, and resistance to systemic therapy in lung cancer.

It has been suggested that concomitant loss of STK11/LKB1 in patients with KRAS mutations may be a unique type of NSCLC, characterized by high infiltration capacity and resistance to therapy [37]. KRAS/LKB1 co-mutation may also be potentially associated with poor ICI treatment effect, STK11/LKB1 activating mutation may be used as a marker (inversely correlated) of NSCLC immunotherapy, predicting immune checkpoint inhibitors (ICIs) ) response, but its predictive role could not be formally determined due to the lack of a sufficient number of ICI-naïve control groups[38].Loss of LKB1 directly or indirectly affects VEGF signaling and may affect other additional angiogenic pathways, thereby promoting angiogenesis[39]. Overall, the role of STK11/LKB1 in lung cancer warrants further investigation.

Two of the eight TTF1/P40 co-expressing NSCLC summarized in this article had NF1 mutations.The NF1 gene encodes neurofibrin, and inactivating mutations of NF1 can lead to a sustained increase in the level of active RAS-GTP in cells and prolonged activation of the RAS/RAF/MAPK signaling pathway, resulting in increased cell proliferation and survival [40]. In addition to small cell lung cancer, NF1 mutations are also found in a variety of tumors, including peripheral nerve sheath tumors, lymphocytic leukemia, urothelial carcinoma, pancreatic cancer, gastric adenocarcinoma, etc[41].

NF1 mutations are found in 7-10% of lung ADCs and approximately 12% of SQCs [42-43]. Investigators have found that NF1-mutant tumors often have a predominantly solid component[44]， and solid content, even in small amounts, is a poor prognostic predictor of lung cancer[45] . About 25% of NF1 mutations co-occur with mutations in known oncogenes KRAS, BRAF, ERBB2, HRAS, and NRAS and rarely co-mutate with other tumor suppressor genes (TP53,RB1 ,STK11/LKB1)[44]. In the case we reported, co-mutation of NF1 and STK11/LKB1 occurred, and the mutation frequency was only 2.5%.There are no other reports of NF1 and STK11/LKB1 co-mutated in lung cancer,more research and analysis are needed for its representative significance.

Five of the eight reported cases showed mutations or polymorphisms in TP53.TP53 mutations are prevalent in patients with NSCLC , with 35% to 55% of patients with NSCLC having TP53 mutations, and are more commonly seen in squamous carcinomas compared to adenocarcinomas[46].TP53 plays an important role in cancer suppression by inducing senescence, blocking the cell cycle and/or apoptosis to regulate cellular responses to various stress signals[47], but both gene mutations and single nucleotide polymorphisms can alter the normal function of p53[48]. Clinically, tumours with TP53 mutations are larger in size, more commonly associated with high clinical staging, lymph node metastasis and worse prognosis, and this gene can be used as a target for molecular therapy and as a prognostic indicator in clinical practice [49].Other mutations were identified in this case, including ARID1A, BRCA2，FGFR1，MSH2，RET and TSC1.Their clinical significance in this case is currently unknown.

The 5th edition of the WHO lung cancer classification does not provide a clear description of NSCLC co-expressing TTF1 and p40 in the same tumour cells as presented here. We prefer to characterize it as a poorly differentiated carcinoma with a dual p40þ/TTF1 phenotype, which may be a new and under-recognized type. It is highly infiltrative, has a poor response to treatment, a poor prognosis and is genetically altered, and more case studies are needed to characterise it.